\documentclass[longauth,letter]{aa}  
\usepackage[english]{babel}
\usepackage[utf8]{inputenc}
\usepackage[T1]{fontenc}
\usepackage{graphicx}
\usepackage{txfonts}
\usepackage{amsmath,amssymb}
\usepackage{etoolbox}
\usepackage{wasysym}
\usepackage{ccaption}

\newcommand{\Ha} {\mbox{H$\alpha$}\,}
\newcommand{\Hb} {\mbox{H$\beta$}\,}
\newcommand{\Hg} {\mbox{H$\gamma$}\,}
\newcommand{\Hd} {\mbox{H$\delta$}\,}
\newcommand{\Hii} {\ion{H}{ii}\,}
\newcommand{\Hei} {\ion{He}{i}\,}
\newcommand{\Caii} {\ion{Ca}{ii}}
\newcommand{\Feii} {\ion{Fe}{ii}}
\newcommand{\Nii} {\ion{N}{ii}}
\newcommand{\Sii} {\ion{S}{ii}}
\newcommand{\Nai} {\ion{Na}{i}\,}

\begin{document} 

\title{SN 2021foa, a transitional event between a Type IIn (SN 2009ip-like) and a Type Ibn supernova}

\titlerunning{The transitional SN 2021foa}

\author{A. Reguitti\inst{1,2,3}\fnmsep\thanks{E-mail: andreareguitti@gmail.com},
A. Pastorello$^{3}$, G. Pignata$^{1,2}$, M. Fraser$^{4}$, M. D. Stritzinger$^{5}$, S. J. Brennan$^{4}$, Y.-Z. Cai$^{6}$,\\
N. Elias-Rosa$^{3,7}$, D. Fugazza$^{8}$, C. P. Gutierrez$^{9,10}$, E. Kankare$^{10,11}$, R. Kotak$^{10}$, P. Lundqvist$^{12,13}$,\\ P. A. Mazzali$^{14,15}$,
S. Moran$^{10}$, I. Salmaso$^{3,16}$, L. Tomasella$^{3}$, G. Valerin$^{3,16}$, H. Kuncarayakti$^{10,9}$
}

\authorrunning{A. Reguitti et al.} 

\institute{Departamento de Ciencias F\'{i}sicas – Universidad Andres Bello, Avda. Rep\'{u}blica 252, 8320000, Santiago, Chile
\and
Millennium Institute of Astrophysics, Nuncio Monsenor S\'{o}tero Sanz 100, Providencia, 8320000, Santiago, Chile
\and
INAF – Osservatorio Astronomico di Padova, Vicolo dell'Osservatorio 5, 35122 Padova, Italy
\and
School of Physics, O’Brien Centre for Science North, University College Dublin, Belfield, Dublin 4, Ireland
\and
Department of Physics and Astronomy, Aarhus University, Ny Munkegade 120, DK-8000 Aarhus C, Denmark
\and
Physics Department and Tsinghua Center for Astrophysics (THCA), Tsinghua University, Beijing 100084, PR China
\and
Institute of Space Sciences (ICE, CSIC), Campus UAB, Carrer de Can Magrans s/n, 08193 Barcelona, Spain
\and
INAF – Osservatorio Astronomico di Brera, via E. Bianchi 46 I-23807, Merate, Italy
\and
Finnish Centre for Astronomy with ESO (FINCA), University of Turku, Vesilinnantie 5, FI-20014, Turku, Finland
\and
Department of Physics and Astronomy, University of Turku, FI-20014 Turku, Finland
\and
Turku Collegium for Science, Medicine and Technology, University of Turku, FI-20014 Turku, Finland
\and
Department of Astronomy, AlbaNova University Center, Stockholm University, SE-10691 Stockholm, Sweden
\and
The Oskar Klein Centre, AlbaNova, SE-10691 Stockholm, Sweden
\and
Astrophysics Research Institute, Liverpool John Moores University, ic2, 146 Brownlow Hill, Liverpool L3 5RF, UK
\and
Max-Planck Institut fur Astrophysik, Karl-Schwarzschild-Str. 1, D-85741 Garching, Germany
\and
Dipartimento di Fisica e Astronomia `G. Galilei’ - Università di Padova, Vicolo dell’Osservatorio 3, 35122 Padova, Italy
}

\date{Received XXX; accepted YYY}
 
\abstract
{We present photometric and spectroscopic data of the unusual interacting supernova (SN) 2021foa. It rose to an absolute magnitude peak of $M_r=-18$ mag in 20 days. The initial light curve decline shows some luminosity fluctuations before a long-lasting flattening. A faint source ($M_r\sim -14$ mag) was detected in the weeks preceding the main event, showing a slow-rising luminosity trend. The $r$-band absolute light curve is very similar to those of SN 2009ip-like events, with a faint and shorter duration brightening (`Event A') followed by a much brighter peak (`Event B').
The early spectra of SN 2021foa show a blue continuum with narrow ($v_{FWHM}\sim$400 km s$^{-1}$) H emission lines, that, two weeks later, reveal a complex profile, with a narrow P Cygni on top of an intermediate-width ($v_{FWHM}\sim$2700 km s$^{-1}$) component. At +12 days metal lines in emission appear, while \Hei lines become very strong, with \Hei~$\lambda$5876 reaching half of the \Ha luminosity, much higher than in previous SN 2009ip-like objects.
We propose SN 2021foa to be a transitional event between the H-rich SN 2009ip-like SNe and the He-rich Type Ibn SNe.}


\maketitle
%

\section{Introduction}

Supernovae (SNe) that explode within a dense and massive circumstellar medium (CSM) are called `interacting' SNe \citep{fraser20}. If the CSM is H-rich, they are Type IIn SNe (\citealt{schlegel}; \citealt{filippenko}), and their spectra show narrow Balmer emission lines. If the CSM is He-rich, they are classified as Type Ibn SNe (\citealt{matheson}; \citealt{pasto08a}; \citealt{hosse17}), and strong emissions from \Hei lines are present.

Among SNe IIn, SN 2009ip (\citealt{pasto13}, \citealt{fraser13,fraser15}, \citealt{margutti}, \citealt{mauerhan14}, \citealt{graham14,graham17}) and similar objects\footnote{SN 2010mc \citep{smp}, SN 2011fh (\citealt{pessi}), LSQ13zm \citep{leo}, SN 2015bh (\citealt{nancy}, \citealt{thone}), SN 2016bdu \citep{pasto18}, AT 2016jbu (\citealt{kilpatrick}, \citealt{brennan1,brennan2}).} are characterized by wide variability or recurrent outbursts in the years prior to the explosion. SN 2009ip has a double-peak light curve with a first luminous ($M_r\sim-15$ mag) maximum just a few weeks before the brightest one ($M_r\sim~-18$ mag). Those peaks are often referred to as `Event A' and `B', respectively \citep{pasto13}.

The second major class of interacting SNe are of Type Ibn. Their light curves usually fade rapidly after peak, and their spectra are dominated by narrow lines of \Hei, and very weak or no H lines. Transitional Type Ibn/IIn SNe showing both H and \Hei lines were also discovered (\citealt{pasto08b,pasto15}; \citealt{smith12}; \citealt{hosse17}), with He lines having a comparable strength as the H ones.

In this paper, we present the photometric and spectroscopic follow-up campaign of SN 2021foa, an interacting SN with a photometric evolution almost identical to SN 2009ip-like objects, with a complex \Ha line profile, but also strong \Hei emission lines.


\section{Discovery and host galaxy} \label{sez2}

SN 2021foa (a.k.a. ASASSN-21dg, ATLAS21htp, PS21cae) was discovered by the All Sky Automated Survey for SuperNovae (ASAS-SN; \citealt{shappee}) on 15~March 2021 (MJD=59288.45) at a Sloan-$g$ apparent magnitude of 15.9, with the last non-detection 10 days earlier, at $g$=17.9 mag \citep{stanek}. Although, ASAS-SN detected it on March 9 at $g$=17.6 mag and observed a 6-days rise to $g$=15.9~mag\footnote{from the ASAS-SN Supernova Patrol}, when the discovery was reported. Its coordinates are $\alpha$=13:17:12.29, $\delta$=$-$17:15:24.19 (J2000). SN 2021foa was classified by \cite{angus}. \
The host galaxy IC 863 is a barred spiral, with a redshift of $z$=0.008386 \citep{pisano}. 
The NASA/IPAC Extragalactic Database (NED)\footnote{https://ned.ipac.caltech.edu/ reports a kinematic distance, corrected for the Virgo Infall, of $d=34.8\pm2.4$ Mpc ($\mu=32.71\pm0.15$ mag), that we adopt as the distance to IC 863.}
The Milky Way reddening towards IC 863 is $A_V$=0.224 mag \citep{schlafly}. From spectroscopic considerations (see Appendix \ref{appendix}) we infer the presence of additional host galaxy extinction of $A_V(host)\approx0.40\pm0.05$ mag.


\section{Photometric evolution}
\label{sez3}

\begin{figure*}
\includegraphics[width=1.02\columnwidth]{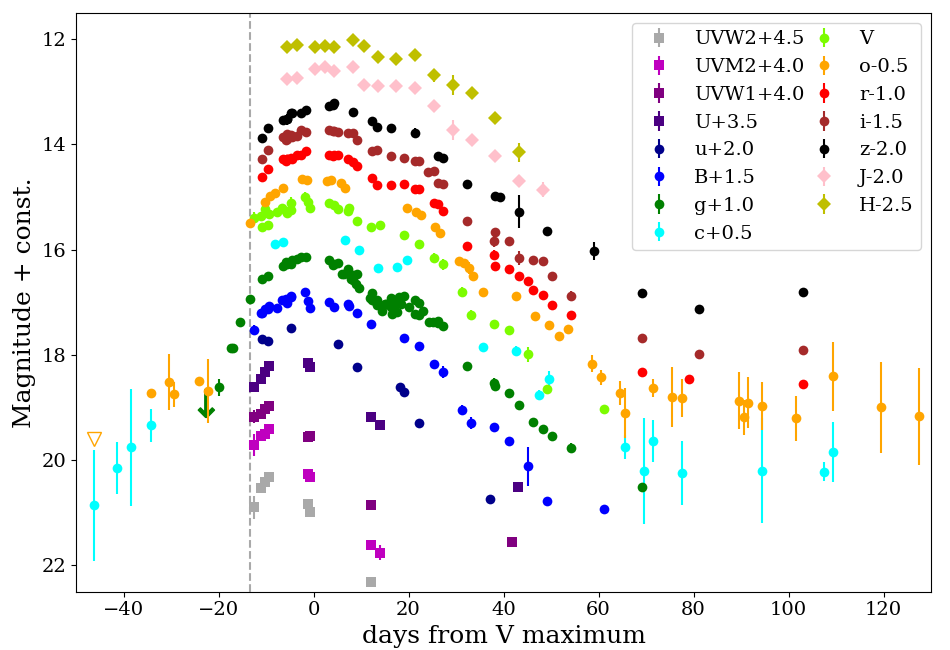} \includegraphics[width=1.02\columnwidth]{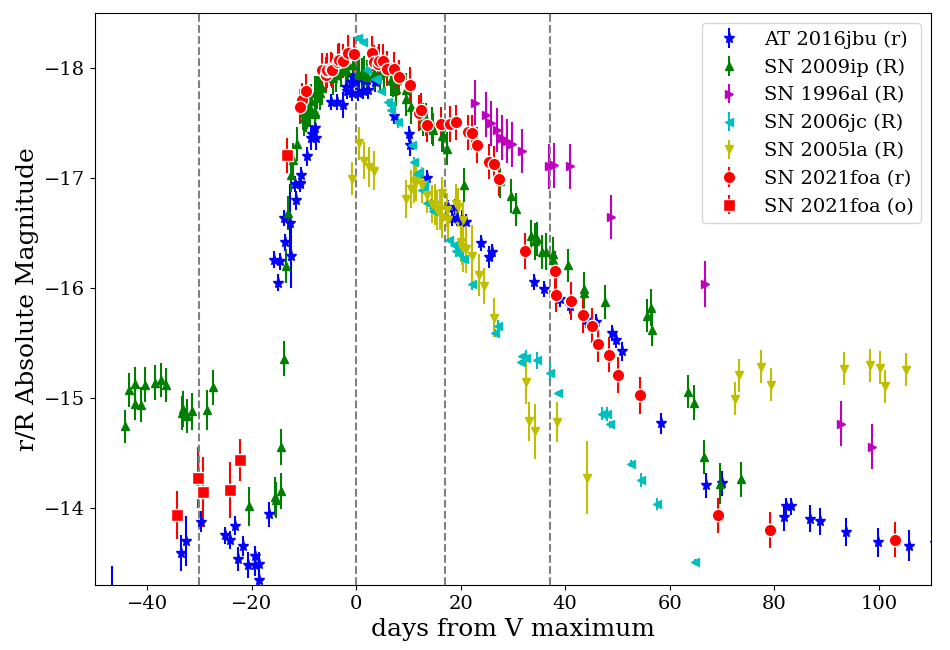}
\caption{Left: UV, Optical and NIR light curves of SN 2021foa, covering 6 months of observations. The phases are relative to the $V$-band maximum. The pre-discovery detections are also reported. The discovery epoch is marked with a vertical line.
Right: Comparison of the $r$-band absolute light curves of SN 2021foa, AT 2016jbu and SN 2009ip ($R$-band).
The error bars include the uncertainties on the photometric measurements, the distance and the reddening.
The vertical lines mark the significant phases: the `\emph{Event A}', the `\emph{Event B}', the `\emph{plateau}' and the `\emph{knee}'.
The ATLAS pre-discovery $orange$-band detections of SN 2021foa are also shown with square symbols to point out its \emph{Event A}.
The $R$-band absolute light curves of the Type IIn SN~1996al, the Type Ibn SN 2006jc and the transitional IIn/Ibn SN 2005la are also plotted.}
\label{fig1}
\end{figure*}

Our multi-band follow-up campaign started soon after discovery, and lasted for 6 months. We collected \textit{Swift} ultraviolet (UV) and ground-based optical/near infrared (NIR) photometric data with a plethora of telescopes and instruments, listed in Table \ref{tab1}.

The optical and NIR photometric data were reduced using standard procedures with the dedicated \textit{Snoopy} pipeline (\citealt{cappellaro}, see \citealt{reguitti} for a description of the procedures). The UV data were reduced with the HEASOFT pipeline\footnote{NASA High Energy Astrophysics Science Archive Research Center - Heasarc 2014.}. The final UV, optical (Sloan, Johnson and ATLAS) and NIR magnitudes are listed in Tables \ref{tab3}-\ref{tab7}, while the light curves are plotted in Fig. \ref{fig1}, left panel.

The CHilean Automatic Supernova sEarch survey (CHASE; \citealt{pignata}) project monitored the field of IC 863 between 2008 and 2015, while the Palomar Transient Factory (PTF; \citealt{law}) scanned it between 2009 and 2014. We inspected their archival images in search of signatures of pre-explosion activity from the progenitor of SN 2021foa, but found no evidence of variability. The Pan-STARRS1 (PS1; \citealt{chambers}) survey also observed the sky region of IC 863 in the years 2013-2020 providing only deep ($\sim$22 mag) upper limits.

The Asteroid Terrestrial-impact Last Alert System (ATLAS, \citealt{tonry}) survey detected the 12 days rise of a faint source (from ATLAS-cyan ($c$) $\sim$20.4 to $\sim$18.8 mag) at the position of SN 2021foa since 10 February 2021, 43 days before the discovery, that remained at nearly constant ATLAS-orange ($o$) $\sim$19~mag for 3 weeks. 

We observed a rise in the first 4 \textit{Swift} epochs. By fitting a 2nd-order polynomial to the data, we found that the UV maximum was reached about 5 days after discovery, while in the optical the peak was reached between 3 and 6 days later (in $B$- and $z$-band), respectively. The $V$-band maximum was reached on MJD=59301.8$\pm$0.1, and we adopt this as a reference epoch. The $V$-band peak absolute magnitude is $M_V=-17.8\pm0.2$ mag.
The light curves are remarkably similar in the different bands: after maximum, the luminosity of the object starts to decline before settling on a plateau (for $\sim$10 days, between +13 and +22 d), about 1 mag fainter than the peak (less in redder bands, e.g. 0.5 mag in $z$). Following the plateau, the light curves displays a rapid and linear decline, lasting $\sim$2 months, with a faster decay in the blue filters when compared to the red ones. The latest observed magnitudes are slightly fainter than the pre-discovery ATLAS ones.
The NIR light curve evolution follows that of the redder optical bands (albeit the NIR campaign lasted only 2 months).
After +80 d, a flattening is observed in the $riz$, $cyan$ and $orange$ light curves. The ATLAS observations continued up to +130 d, when it was stopped as the object was too close to the Sun.

As shown in Fig. \ref{fig1}, right panel, we found a remarkable similarity among the $r$-band absolute light curves of SN 2021foa with SN 2009ip during the brightest event (\citealt{pasto13}, \citealt{fraser13}), and also with the SN 2009ip-like object AT 2016jbu (\citealt{kilpatrick}, \citealt{brennan1}). Comparing them with the H-rich Type IIn SN 1996al \citep{benetti16}, the He-rich SN 2006jc \citep{pasto07} and the transitional IIn/Ibn SN 2005la \citep{pasto08b}, we see that the decline rate of SN 2021foa is intermediate between them.
The faint ATLAS pre-discovery detections (at $M_o\sim-14$ mag) correspond to the \emph{Event A}, while the brighter post-discovery light curve peak is the \emph{Event B}. The \emph{plateau} at +20 d is more pronounced in SN 2021foa, while the \emph{knee} occurs slightly earlier (+40 d instead of 45 d) and is less noticeable. Finally, both AT 2016jbu and SN 2021foa show a much slower decline in their light curves roughly from +70 d onwards.


\section{Spectral evolution}
\label{sez4}

\begin{figure*}
\includegraphics[width=1.5\columnwidth]{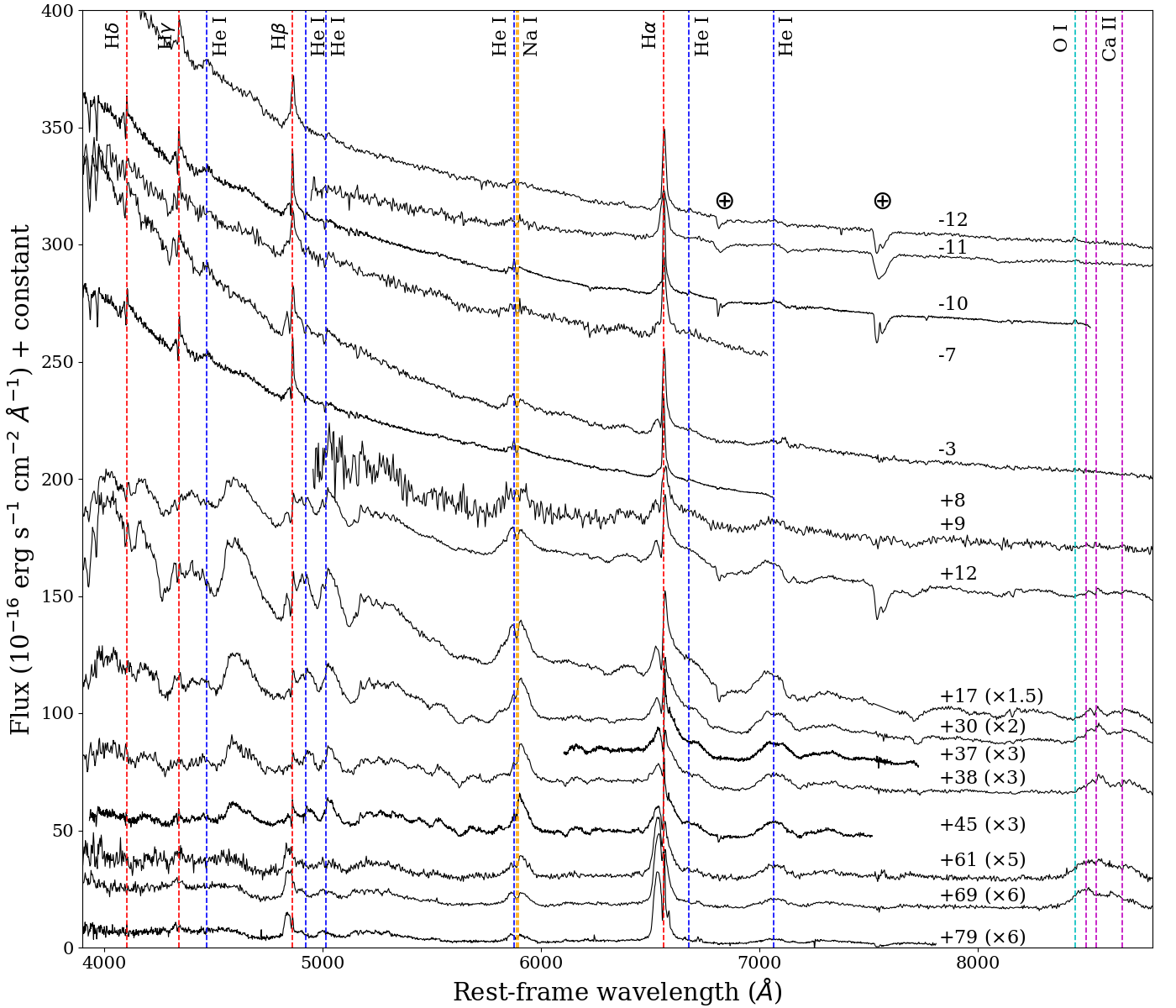}\includegraphics[width=0.54\columnwidth]{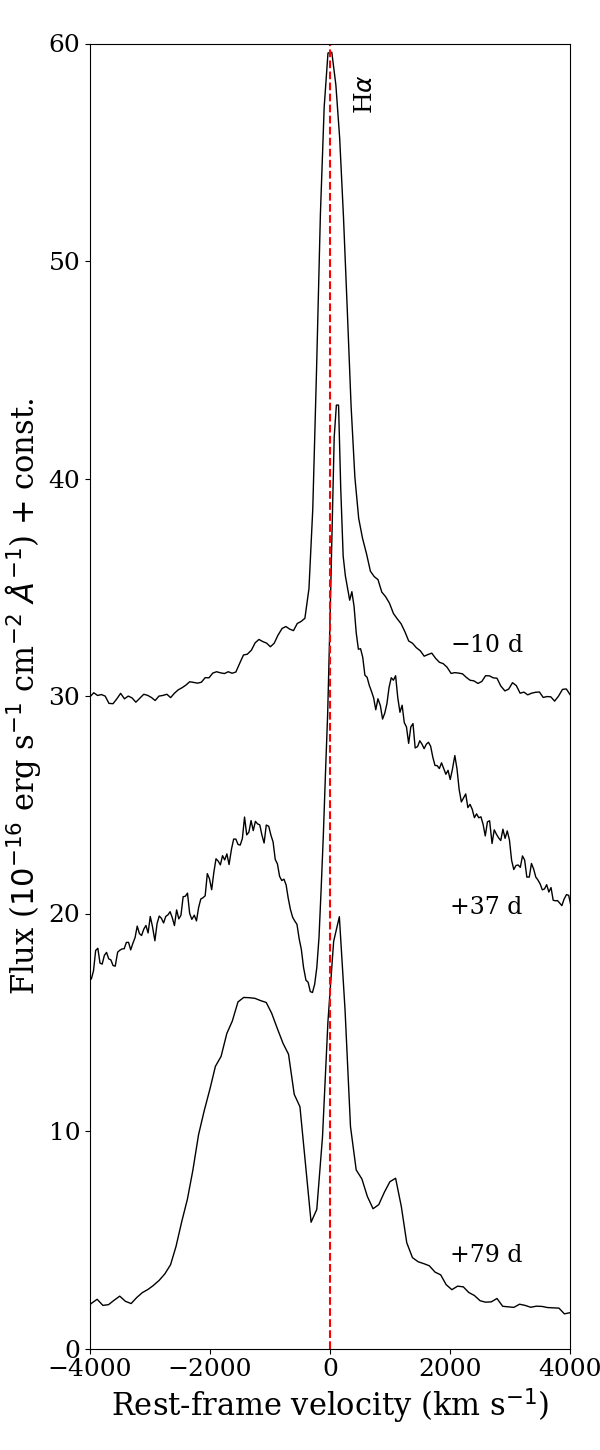}
\caption{Left: sequence of spectra of SN 2021foa. The spectra are redshift- and reddening-corrected. The principal identified lines are marked, as well as the telluric absorption bands. The phases indicated are relative to the $V$-band maximum. For a better visualization, the fluxes of the late spectra are multiplied by a factor reported in parenthesis. Right: zoom on the \Ha line at three representative epochs: $-$10 d, +37 d and +79 d, to highlight the evolution and the complexity of its profile. In abscissa are the rest-frame velocities.}
\label{fig2}
\end{figure*}

We conducted also a spectroscopic follow-up of SN 2021foa, during which we collected 19 optical spectra that span the first 3 months of evolution. The log of spectroscopic observations is provided in Table \ref{tab2}, and the time-series is presented in Fig. \ref{fig2}, left panel.
The spectra from `Copernico', Nordic Optical Telescope (NOT) and Gran Telescopio Canarias (GTC) telescopes were reduced, extracted and calibrated with slightly different versions of the \texttt{Foscgui}\footnote{https://sngroup.oapd.inaf.it/foscgui.html} pipeline \citep{cappellaro}, adapted for the AFOSC, ALFOSC and OSIRIS instruments. The spectra from the Telescopio Nazionale Galileo (TNG) were reduced with the standard procedures under the \texttt{PyRAF} environment.
The final spectra were fine-tuned against the closest photometry.

The early spectra show a blue and hot continuum with a black-body temperature $T_{BB}$ of 15000 K, and narrow H lines in emission, as typically observed in SNe IIn. \Hei lines ($\lambda$4471, $\lambda$5015 and $\lambda$5876) are present but weak. 
Two days later (at $-$10~d), we took a mid-resolution spectrum. In this spectrum the \Hb, \Hg and \Hd lines start to develop a narrow P Cygni absorption on top of the intermediate component. From \Hb, the position of the P Cygni minimum corresponds to an expansion velocity of 500 km s$^{-1}$. 
We deconvolved the \Ha profile into two Gaussian components: a narrow component with a full-width at half maximum (FHWM) velocity\footnote{corrected for instrumental resolution, i.e. $FWHM_{corrected}=\sqrt{FWHM_{observed}^2-FWHM_{instrument}^2}$} of 400 km s$^{-1}$ and an intermediate one of 2700 km s$^{-1}$.

From the 3 days before $V$-maximum spectrum onwards, \Ha starts to show a narrow P Cygni absorption profile. At this epoch, $T_{BB}$ has dropped to 12000 K. 
Two weeks after maximum, the spectrum dramatically changes: in the blue part, emission lines from metals appear, mostly with P Cygni profiles (such as the multiplet 42 of Fe II $\lambda\lambda$4924, 5018 and 5169), while \Hei lines are now very strong, particularly $\lambda$5876, with a flux that is close to half of that of \Ha. A~deep absorption feature is visible on top of the \Hei $\lambda$5876 line. The Balmer lines also reveal P Cygni absorptions, up to H$\varepsilon$. The narrow P Cygni absorption of \Ha is now evident, and the intermediate-width component has turned into a broad one, with a FWHM velocity ($v_{\mathrm{FWHM}}$) of $\sim$8000 km s$^{-1}$. 

Later, the P Cygni profiles tend to disappear (except \Ha), the metal lines broaden, while the \Hei lines remain strong (at +12 d, we measure the following flux ratios: \Hei $\lambda$5876/H$\alpha\simeq$1/2 and \Hei $\lambda$7065/H$\alpha\simeq$1/4). 
At +30 d, the lines become more prominent relatively to the continuum and broaden, with a mean $v_{\mathrm{FWHM}}$ of $\sim$5000 km s$^{-1}$. The P~Cygni absorption in \Ha is less evident, while \Hb weakens. 
The \Hei lines are still more prominent than most Balmer lines, and a broad double-peak bump from the \Caii\,II NIR triplet emerges, as well as a feature around 7300 \AA~that can be attributed to [\Caii] $\lambda\lambda$7291,7324, as its profile is comparable to that of the Ca II NIR triplet, or alternatively \Hei $\lambda$7281. 
Furthermore, a broad and strong emission centered at 4600~\AA~is present in the blue part, possibly due to \Feii. The temperature has cooled to $T\sim$7000 K, based on the peak of the continuum flux. 

In the +37 d mid-resolution spectrum, we deconvolved \Ha in a broad ($v_{\mathrm{FWHM}}\sim$6000 km s$^{-1}$) and narrow ($v_{\mathrm{FWHM}}\approx$450 km s$^{-1}$) emissions, and a narrow P Cygni absorption, with a velocity at the minimum position that is consistent with the FWHM of the narrow emission component. A red shoulder of \Ha is consistent with the emerging \Hei $\lambda$6678 line. The \Hei $\lambda$7065 line has a trapezoidal shape, with $v_{\mathrm{FWHM}}\sim$6000 km s$^{-1}$.
At about 2 months after maximum, the \Hei lines weaken, with \Hei $\lambda$5876/H$\alpha\simeq$ 1/3, and the P Cygni absorption on top of \Ha becomes less pronounced.


\section{Discussion and conclusion}
\label{sez5}
\begin{figure}
\includegraphics[width=1.02\columnwidth]{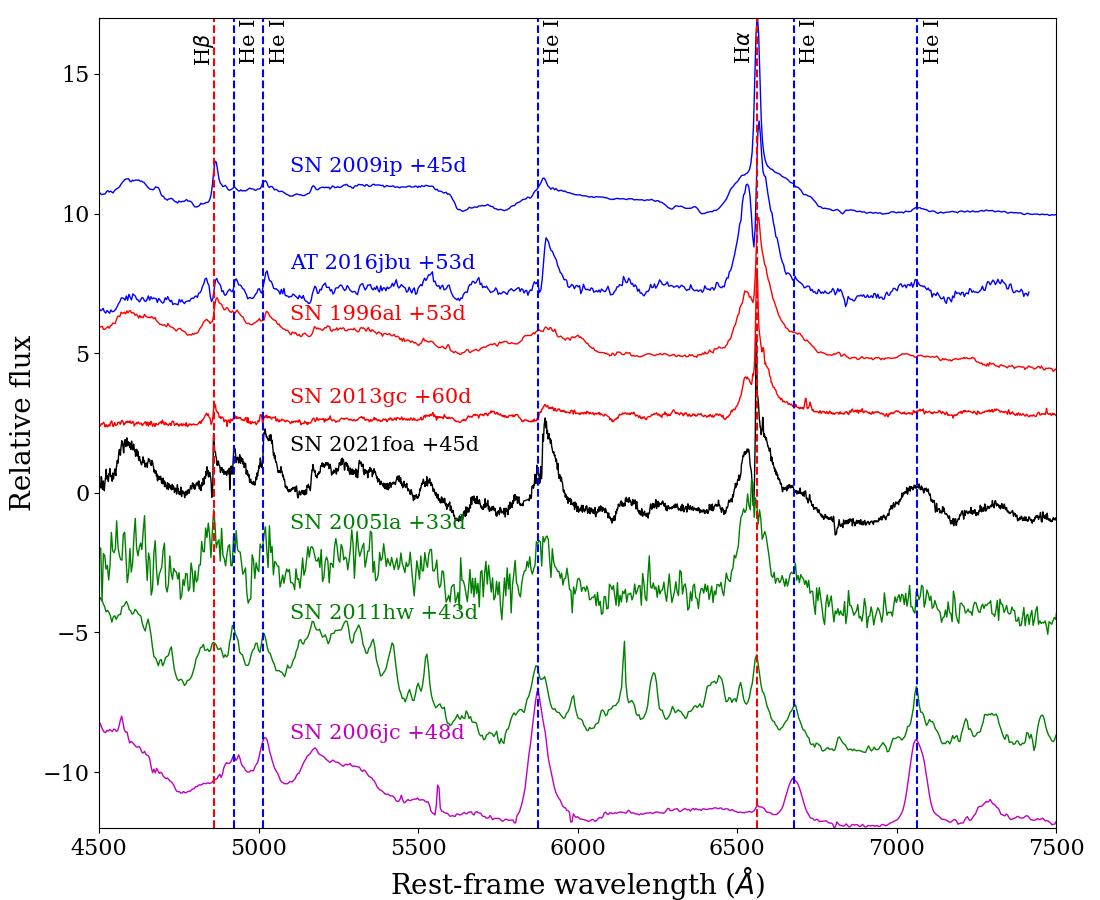}
\caption{Spectral comparison at a similar phase (around 1.5 months after $V$-band maximum) of SN 2021foa, Type IId SNe 2013gc and 1996al, SN 2009ip and the 2009ip-like event AT~2016jbu, the prototypical Type Ibn SN 2006jc, and the transitional Type Ibn/IIn SNe 2005la and 2011hw. Different colours indicate different SN types: SN 2009ip-like in blue, SNe IId in red, transitional IIn/Ibn in green and Ibn in purple. }
\label{fig3}
\end{figure}

The complex Balmer emission line profiles in SN 2021foa, especially \Ha (Fig. \ref{fig2}, right panel), with the simultaneous presence of multiple emission components and a narrow P Cygni absorption, are a distinctive characteristic of a subclass of SNe IIn sometimes labelled as SNe IId (\citealt{benetti00}, \citealt{benetti16}, \citealt{reguitti}). SN 2009ip-like events also reveal a similar structured profile, though they do not show strong \Hei lines. In Fig. \ref{fig3}, we compare the spectral region 4500-7500 \AA~of SN 2021foa, two SNe IId (SNe 2013gc and 1996al), SN 2009ip \citep{pasto13} and AT 2016jbu \citep{brennan1} at about 1.5 months after $V$-band maximum.
We note that the \Ha profiles are quite similar, while the He~I lines in SN 2021foa are much stronger than the comparison objects.

Conversely, the blue part of the spectra of SN 2021foa and the strength of the \Hei lines resemble those of He-rich Type Ibn SNe (\citealt{pasto07}, \citealt{hosse17}). Two notable objects are SN 2011hw (\citealt{smith12}, \citealt{pasto15}), a transitional SN Ibn/IIn that shows an \Ha emission in an otherwise He-dominated spectrum, and SN 2005la \citep{pasto08b}, in which \Ha is even stronger than the He lines. In both objects, an Ofpe/WN9, in transition from a Luminous Blue Variable (LBV) to an early H-poor but not H-free Wolf-Rayet (WR) star, was suggested as a progenitor. In the lower part of Fig.~\ref{fig3}, SN 2021foa is compared with the Type Ibn SN~2006jc \citep{pasto07} and the transitional IIn/Ibn SNe 2005la and 2011hw. Those objects show a progressive strengthening of the \Ha emission, while the \Hei lines remain prominent.
SN~2021foa may be part of a bridge connecting H-rich SN 2009ip-like and Type Ibn SN events, indicating the possible existence of a continuum in properties, mass-loss history and progenitor types between these two types of peculiar transients. The host galaxy metallicity plays an important role on the mass loss history of massive stars, as a metal-poor environment is expected to inhibit mass loss in massive stars, in contrast with what happens with metal rich environments. Indeed, the metallicity near the site of SN 2021foa (see Appendix \ref{appendix}) is roughly Solar.

The outer envelope of the progenitor of SN 2021foa was still H-rich, as at late phases \Ha emission remains the predominant spectral feature, but a larger fraction was lost with respect to SN~2009ip. 
The suggested progenitors of SN 2009ip-like events are H-rich LBV stars (\citealt{smith10}, \citealt{foley11}, \citealt{mauerhan13}, \citealt{smp}, but see \citealt{brennan2}). SNe IId are probably connected to those objects by having similar progenitors, but with a different mass-loss history or observed with a different orientation.
The supposed progenitors of SNe Ibn are H-poor WR stars \citep{foley07}, but \cite{sun} concluded that SNe Ibn can originate from lower-mass stars ($M<$12 $M_{\odot}$) in interacting binaries. The detonation of a Helium white dwarf scenario was also proposed (\citealt{sanders}; \citealt{hosse19}). As SN 2021foa shares photometric and spectroscopic properties with SN 2009ip and SNe IId, but with strong \Hei lines that resemble the spectra of IIn/Ibn SNe, the progenitor of SN 2021foa could have been an LBV on the way becoming a WR star. The star has likely lost a large fraction of its H envelope, although a residual H layer is still retained. 
The wind velocity derived from \Ha ($\sim$450 km s$^{-1}$) is relatively low for a classical WR. While is consistent with the wind velocity from an LBV (e.g. \citealt{vink}), it is also compatible with the wind from a WNH star \citep{smith16}, and is similar to that observed for SN 2005la \citep{pasto08b}.

The upcoming 10 year Legacy Survey of Space and Time (LSST) at the Vera Rubin Telescope will discover hundreds of transitional objects. Statistical studies of transients similar to SN~2021foa and their environments will enable us to elucidate their uncertain nature.


\begin{acknowledgements}
\begin{small}
We gratefully acknowledge the anonymous referee for his/her thorough review of the manuscript.
AR acknowledges support from ANID BECAS/DOCTORADO NACIONAL 21202412. 
MDS is supported by grants from the VILLUM FONDEN (grant no. 28021) and the Independent Research Fund Denmark (IRFD; 8021-00170B). 
NER acknowledges partial support from MIUR, PRIN 2017 (grant 20179ZF5KS), from the Spanish MICINN grant PID2019-108709GB-I00 and FEDER funds, and from the program Unidad de Excelencia María de Maeztu CEX2020-001058-M.
YZC is funded by China Postdoctoral Science Foundation (grant no. 2021M691821). 
SJB would like to thank their support from Science Foundation Ireland and the Royal Society (RS-EA/3471).
HK was funded by the Academy of Finland projects 324504 and 328898.
GP acknowledges support to ANID – Millennium Science Initiative – ICN12\_009.
Based on observations collected at Copernico and Schmidt telescopes (Asiago, Italy) of the INAF - Osservatorio Astronomico di Padova.
Based on observations collected with the Rapid Eye Mount telescope of the Instituto Nazionale di Astrofisica (INAF), hosted at the ESO La Silla Observatory, under programs ID 42208 and 43308.
Based on observations made with the Nordic Optical Telescope, owned in collaboration by the University of Turku and Aarhus University, and operated jointly by Aarhus University, the University of Turku and the University of Oslo, representing Denmark, Finland and Norway, the University of Iceland and Stockholm University at the Observatorio del Roque de los Muchachos, La Palma, Spain, of the Instituto de Astrofisica de Canarias.
Based on observations made with the Liverpool Telescope, operated by Liverpool John Moores University, with financial support from the UK Science and Technology Facilities Council, at the Spanish Observatorio del Roque de los Muchachos, La Palma, Spain, of the Instituto de Astrofisica de Canarias.
The NUTS program is funded in part by the IDA (Instrument Centre for Danish Astronomy).
Based on observations made with the Gran Telescopio Canarias, installed at the Spanish Observatorio del Roque de los Muchachos of the Instituto de Astrofísica de Canarias, in the island of La Palma.
We acknowledge the use of public data from the Swift data archive.
The ATLAS project is primarily funded through NASA grants NN12AR55G, 80NSSC18K0284 and 80NSSC18K1575.
ASAS-SN is supported by the Gordon and Betty Moore Foundation through grant GBMF5490 to the Ohio State University and NSF grant AST-1515927.
\end{small}
\end{acknowledgements}

\bibliographystyle{aa} 


\begin{appendix} 
\section{Host galaxy metallicity and reddening}
\label{appendix}

From our longslit spectroscopy of SN 2021foa at late phase, we extracted the spectrum of an \Hii region adjacent to the SN. The spectrum shows typical narrow emission lines from ionised gas, including \Ha, [\Nii], and [\Sii]. 
Assuming that the adjacent \Hii region is representative for the SN explosion site, we measured the emission line flux of \Ha and [\Nii] $\lambda$6584 in its spectrum to derive the oxygen abundance as a metallicity proxy, via the N2 index according to the \cite{marino} calibration, and also the [\Sii] $\lambda$6717,6731 doublet for the same purpose using the \cite{dopita} scale (D16). The measured metallicity in 12+log(O/H) is 8.59 dex (N2) and 8.66 dex (D16). Within the typical metallicity calibration error of 0.1-0.2 dex, these values agree with each other. The derived metallicity is thus consistent with being nearly solar (12+log(O/H)$_\odot$ = 8.69 dex, \citealt{asplund}).

The estimate of the line of sight reddening is a crucial step for the characterization of a stellar transient. One of the most popular tools is through the detection of narrow interstellar lines. \cite{turatto} proposed to infer the colour excess using a linear relation with the equivalent width (EW) of the \Nai $\lambda\lambda$5890,5896 doublet. \cite{poznanski} revised the relation using the individual line components in higher resolution spectra. \\ In our early spectra of SN 2021foa a narrow absorption of the \Nai~doublet is visible on top of the \Hei $\lambda$5876 line at the host galaxy redshift, with EW=0.8$\pm$0.1 \AA. The \cite{poznanski} relation between sodium absorption and dust extinction saturates at equivalent widths beyond 0.8 \AA, hence we estimate the internal extinction using the \cite{turatto} formula, that provides an additional reddening of $A_V(host)\approx0.40\pm0.05$ mag.

\begin{figure}
\includegraphics[width=1.02\columnwidth]{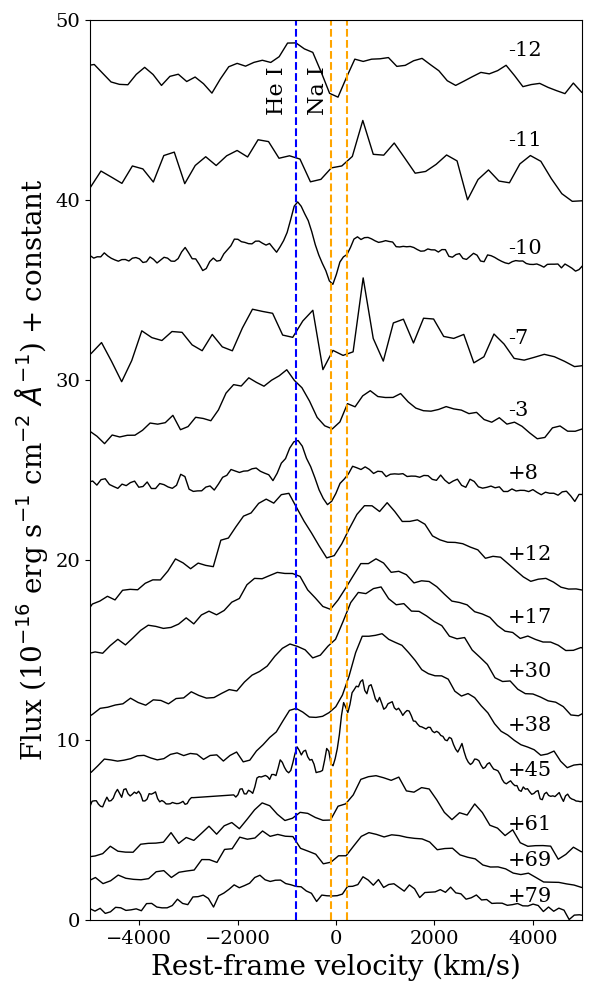}
\caption{Evolution of the profile of the \Nai~absorption with time, in velocity space.}
\label{fig4}
\end{figure}

In Fig. \ref{fig4}, we show the evolution of the profile of the narrow (interstellar) \Nai D feature in the velocity space. While its EW remains roughly constant (within the measurement errors) until +45 d, it seems to significantly increase in the late-time spectra. However, the change of the relative intensities of the broader features of \Hei $\lambda$5876 and \Nai D (attributed to the SN ejected material) affects a reliable estimate of the EW of the interstellar \Nai D component, and probably explains its apparent evolution without the need of invoking changes in the ionization state of the ISM.
As a consequence, in this paper we assume that the EW of the \Nai D absorption measured in the early spectra is entirely produced by interstellar gas, and can be used as a proxy for estimating the reddening contribution of the host galaxy.


\section{Tables} 

\begin{table}
\caption{Observational facilities and instrumentation used in the photometric follow-up of SN 2021foa.}
\label{tab1}
\begin{tabular}{llll}
\hline
Telescope & Location & Instrument & Filters \\
\hline
\textit{Swift} (0.3m) & Space & UVOT & $UV$ filters+$UBV$ \\
\hline
ASAS-SN (0.14m) & Texas & ``Leavitt" & $g$ \\
PROMPT (0.4m+0.6m) & CTIO & Apogee & $BVgriz$ \\
ATLAS (0.5m) & Hawaii & ACAM1 & $c,o$ \\
REM (0.6m) & La Silla & ROS2 & $griz$ \\
Schmidt (0.67m) & Asiago & Moravian & $uBVgri$ \\
Copernico (1.82m) & Asiago & AFOSC & $iz$ \\
LT (2.0m) & La Palma & IO:O & $uBVgriz$ \\
NOT (2.56m) & La Palma & ALFOSC & $uBVgriz$ \\
\hline
REM (0.5m) & La Silla & REMIR & $JH$ \\
\hline
\end{tabular}
\end{table}

\begin{table}
\caption{\textit{Swift} UV VEGA magnitudes of SN 2021foa. All measuremenets are from the UVOT instrument.}
\label{tab3}
\begin{tabular}{ccccc}
\hline
Date & MJD & $UVW2$ & $UVM2$ & $UVW1$ \\
\hline
2021-03-16	&	59289.21	&	16.40$\pm$0.22	&	15.72$\pm$0.21	&	15.18$\pm$0.12	\\
2021-03-17	&	59290.73	&	16.03$\pm$0.04	&	15.54$\pm$0.03	&	15.12$\pm$0.03	\\
2021-03-18	&	59291.64	&	15.91$\pm$0.04	&	15.50$\pm$0.03	&	15.03$\pm$0.03	\\
2021-03-19	&	59292.44	&	15.82$\pm$0.03	&	15.41$\pm$0.03	&	14.98$\pm$0.03	\\
2021-03-27	&	59300.61	&	16.33$\pm$0.05	&	16.27$\pm$0.05	&	15.57$\pm$0.05	\\
2021-03-28	&	59301.15	&	16.49$\pm$0.06	&	16.32$\pm$0.06	&	15.55$\pm$0.05	\\
2021-04-09	&	59313.89	&	17.82$\pm$0.09	&	17.61$\pm$0.08	&	16.86$\pm$0.07	\\
2021-04-11	&	59315.74	&	-	&	17.76$\pm$0.14	&	-	\\
2021-05-09	&	59343.68	&	-	&	-	&	17.55$\pm$0.04 	\\
\hline
\end{tabular}
\end{table}

\begin{table*}
\centering
\caption{Johnson $UBV$ VEGA magnitudes of SN 2021foa.}
\label{tab4}
\begin{tabular}{cccccl}
\hline
Date & MJD & $U$ & $B$ & $V$ & Instrument \\
\hline
2021-03-16	&	59289.21	&	15.12$\pm$0.08	&	16.02$\pm$0.09	&	15.40$\pm$0.11	&	UVOT \\
2021-03-17	&	59290.73	&	14.97$\pm$0.06	&	15.71$\pm$0.08	&	15.37$\pm$0.05	&	UVOT \\
2021-03-17	&	59290.99	&	-	&	15.71$\pm$0.02	&	15.56$\pm$0.03	&	Moravian \\
2021-03-18	&	59291.64	&	14.83$\pm$0.06	&	15.64$\pm$0.05	&	15.23$\pm$0.05	&	UVOT \\
2021-03-19	&	59292.12	&	-	&	15.63$\pm$0.01	&	15.53$\pm$0.01	& IO:O \\
2021-03-19	&	59292.44	&	14.70$\pm$0.05	&	15.58$\pm$0.03	&	15.32$\pm$0.04	&	UVOT \\
2021-03-21	&	59294.13	&	-	&	15.61$\pm$0.08	&	15.29$\pm$0.04	& Apogee \\
2021-03-22	&	59295.13	&	-	&	15.46$\pm$0.04	&	15.21$\pm$0.08	& Apogee \\
2021-03-22	&	59295.97	&	-	&	15.44$\pm$0.05	&	15.26$\pm$0.03	&	Moravian \\
2021-03-23	&	59296.12	&	-	&	15.52$\pm$0.05	&	15.30$\pm$0.05	& Apogee \\
2021-03-24	&	59297.06	&	-	&	15.38$\pm$0.01	&	15.25$\pm$0.01	& IO:O \\
2021-03-24	&	59297.12	&	-	&	15.39$\pm$0.06	&	15.11$\pm$0.11	& Apogee \\
2021-03-27	&	59300.11	&	-	&	15.30$\pm$0.07	&	14.99$\pm$0.08	& Apogee \\
2021-03-27	&	59300.61	&	14.65$\pm$0.04	&	15.48$\pm$0.04	&	15.10$\pm$0.05	&	UVOT \\
2021-03-28	&	59301.15	&	14.74$\pm$0.04	&	15.60$\pm$0.04	&	15.21$\pm$0.06	&	UVOT \\
2021-04-01	&	59305.09	&	-	&	15.50$\pm$0.04	&	15.11$\pm$0.05	& Apogee \\
2021-04-02	&	59306.10	&	-	&	15.58$\pm$0.04	&	15.15$\pm$0.06	& Apogee \\
2021-04-02	&	59306.95	&	-	&	-	&	15.22$\pm$0.02	&	Moravian \\
2021-04-04	&	59308.98	&	-	&	15.53$\pm$0.02	&	15.27$\pm$0.01	&	Moravian \\
2021-04-05	&	59309.17	&	-	&	15.56$\pm$0.06	&	15.21$\pm$0.06	& Apogee \\
2021-04-07	&	59311.02	&	-	&	15.70$\pm$0.02	&	15.46$\pm$0.02	&	Moravian \\
2021-04-09	&	59313.89	&	15.68$\pm$0.09	&	15.91$\pm$0.04	&	15.57$\pm$0.06	&	UVOT \\
2021-04-11	&	59315.74	&	15.83$\pm$0.09	&	-	&	15.53$\pm$0.06	&	UVOT \\
2021-04-16	&	59320.95	&	-	&	16.19$\pm$0.02	&	15.72$\pm$0.03	&	Moravian \\
2021-04-19	&	59323.98	&	-	&	16.33$\pm$0.02	&	15.89$\pm$0.04	&	Moravian \\
2021-04-23	&	59327.13	&	-	&	16.68$\pm$0.08	&	16.16$\pm$0.08	& Apogee \\
2021-04-25	&	59329.04	&	-	&	16.83$\pm$0.11	&	16.28$\pm$0.10	& Apogee \\
2021-04-29	&	59333.02	&	-	&	17.55$\pm$0.09	&	16.80$\pm$0.08	& Apogee \\
2021-05-01	&	59335.01	&	-	&	17.79$\pm$0.11	&	17.24$\pm$0.09	& Apogee \\
2021-05-05	&	59339.89	&	-	&	17.88$\pm$0.07	&	17.42$\pm$0.08	&	Moravian \\
2021-05-08	&	59342.86	&	-	&	18.14$\pm$0.05	&	17.52$\pm$0.05	&	Moravian \\
2021-05-10	&	59344.75	&	17.00$\pm$0.03	&	-	&	-	&	UVOT \\
2021-05-13	&	59347.03	&	-	&	18.62$\pm$0.37	&	17.99$\pm$0.14	& Apogee \\
2021-05-16	&	59350.90	&	-	&	19.27$\pm$0.04	&	18.65$\pm$0.02	& IO:O \\
2021-05-29	&	59363.01	&	-	&	19.43$\pm$0.03	&	19.02$\pm$0.03	& ALFOSC \\
\hline
\end{tabular}
\end{table*}

\begin{table*}
\centering
\caption{Sloan $ugriz$ AB magnitudes of SN 2021foa.}
\label{tab5}
\begin{tabular}{cccccccl}
\hline
Date & MJD & $u$ & $g$ & $r$ & $i$ & $z$ & Instrument \\
\hline
2021-03-05	&	59278.41	&	-	&	>17.9			&	-	&	-	&	-	&	Leavitt	\\
2021-03-09	&	59282.01	&	-	&	17.62$\pm$0.16	&	-	&	-	&	-	&	Leavitt	\\
2021-03-11	&	59284.87	&	-	&	16.87$\pm$0.07	&	-	&	-	&	-	&	Leavitt	\\
2021-03-11	&	59284.37	&	-	&	16.87$\pm$0.07	&	-	&	-	&	-	&	Leavitt	\\
2021-03-13	&	59286.31	&	-	&	16.38$\pm$0.05	&	-	&	-	&	-	&	Leavitt	\\
2021-03-15	&	59288.45	&	-	&	15.93$\pm$0.05	&	-	&	-	&	-	&	Leavitt	\\
2021-03-17	&	59290.99	&	15.70$\pm$0.02	&	15.55$\pm$0.02	&	15.61$\pm$0.03	&	15.77$\pm$0.03	&	-	&	Moravian	\\
2021-03-18	&	59291.05	&	-	&	-	&	-	&	-	&	15.87$\pm$0.04	&	AFOSC	\\
2021-03-19	&	59292.12	&	15.74$\pm$0.01	&	15.51$\pm$0.01	&	15.47$\pm$0.01	&	15.61$\pm$0.01	&	15.70$\pm$0.01	&	IO:O	\\
2021-03-22	&	59295.31	&	-	&	15.32$\pm$0.02	&	15.28$\pm$0.03	&	15.36$\pm$0.03	&	15.53$\pm$0.05	&	Apogee	\\
2021-03-22	&	59295.96	&	-	&	15.24$\pm$0.04	&	15.32$\pm$0.05	&	15.41$\pm$0.04	&	-	&	Moravian	\\
2021-03-23	&	59296.22	&	-	&	15.27$\pm$0.02	&	15.28$\pm$0.02	&	15.40$\pm$0.02	&	15.52$\pm$0.03	&	Apogee	\\
2021-03-23	&	59296.23	&	-	&	15.24$\pm$0.02	&	15.28$\pm$0.01	&	15.31$\pm$0.06	&	15.52$\pm$0.10	&	ROS2	\\
2021-03-24	&	59297.06	&	15.49$\pm$0.01	&	15.27$\pm$0.01	&	15.27$\pm$0.01	&	15.32$\pm$0.01	&	15.39$\pm$0.01	&	IO:O	\\
2021-03-24	&	59297.18	&	-	&	15.19$\pm$0.02	&	15.28$\pm$0.02	&	15.35$\pm$0.03	&	15.41$\pm$0.03	&	Apogee	\\
2021-03-25	&	59298.24	&	-	&	15.18$\pm$0.03	&	15.20$\pm$0.04	&	15.33$\pm$0.05	&	-	&	ROS2	\\
2021-03-26	&	59299.23	&	-	&	15.14$\pm$0.03	&	15.19$\pm$0.02	&	15.22$\pm$0.03	&	15.40$\pm$0.03	&	ROS2	\\
2021-03-27	&	59300.14	&	-	&	15.13$\pm$0.02	&	15.12$\pm$0.02	&	15.25$\pm$0.03	&	15.34$\pm$0.04	&	Apogee	\\
2021-04-01	&	59305.16	&	-	&	15.20$\pm$0.01	&	15.21$\pm$0.02	&	15.23$\pm$0.02	&	15.28$\pm$0.03	&	Apogee	\\
2021-04-02	&	59306.01	&	-	&	15.28$\pm$0.02	&	15.22$\pm$0.01	&	15.24$\pm$0.03	&	15.25$\pm$0.04	&	ROS2	\\
2021-04-02	&	59306.08	&	-	&	15.28$\pm$0.02	&	15.19$\pm$0.01	&	15.24$\pm$0.02	&	15.20$\pm$0.02	&	Apogee	\\
2021-04-02	&	59306.94	&	15.80$\pm$0.02	&	15.26$\pm$0.02	&	15.20$\pm$0.03	&	15.26$\pm$0.03	&	-	&	Moravian	\\
2021-04-03	&	59307.86	&	-	&	15.46$\pm$0.03	&	-	&	-	&	-	&	Leavitt	\\
2021-04-04	&	59308.20	&	-	&	15.46$\pm$0.05	&	-	&	-	&	-	&	Leavitt	\\
2021-04-04	&	59308.98	&	-	&	15.37$\pm$0.02	&	15.27$\pm$0.01	&	15.29$\pm$0.02	&	-	&	Moravian	\\
2021-04-05	&	59309.78	&	-	&	15.55$\pm$0.03	&	-	&	-	&	-	&	Leavitt	\\
2021-04-06	&	59310.06	&	-	&	15.48$\pm$0.03	&	15.34$\pm$0.03	&	15.28$\pm$0.02	&	15.39$\pm$0.05	&	ROS2	\\
2021-04-06	&	59310.08	&	-	&	15.58$\pm$0.03	&	-	&	-	&	-	&	Leavitt	\\
2021-04-06	&	59310.78	&	-	&	15.65$\pm$0.03	&	-	&	-	&	-	&	Leavitt	\\
2021-04-07	&	59311.02	&	16.24$\pm$0.03	&	15.47$\pm$0.01	&	15.41$\pm$0.02	&	15.42$\pm$0.03	&	-	&	Moravian	\\
2021-04-07	&	59311.31	&	-	&	15.73$\pm$0.04	&	-	&	-	&	-	&	Leavitt	\\
2021-04-09	&	59313.76	&	-	&	15.92$\pm$0.04	&	-	&	-	&	-	&	Leavitt	\\
2021-04-10	&	59314.09	&	-	&	15.96$\pm$0.04	&	-	&	-	&	-	&	Leavitt	\\
2021-04-10	&	59314.17	&	-	&	15.82$\pm$0.02	&	15.64$\pm$0.02	&	15.63$\pm$0.02	&	15.55$\pm$0.03	&	Apogee	\\
2021-04-10	&	59314.92	&	-	&	16.00$\pm$0.04	&	-	&	-	&	-	&	Leavitt	\\
2021-04-11	&	59315.22	&	-	&	16.06$\pm$0.04	&	-	&	-	&	-	&	Leavitt	\\
2021-04-11	&	59315.27	&	-	&	15.96$\pm$0.03	&	15.77$\pm$0.03	&	15.61$\pm$0.04	&	15.67$\pm$0.07	&	ROS2	\\
2021-04-12	&	59316.13	&	-	&	16.18$\pm$0.04	&	-	&	-	&	-	&	Leavitt	\\
2021-04-12	&	59316.79	&	-	&	16.03$\pm$0.05	&	-	&	-	&	-	&	Leavitt	\\
2021-04-13	&	59317.12	&	-	&	16.03$\pm$0.04	&	-	&	-	&	-	&	Leavitt	\\
2021-04-13	&	59317.81	&	-	&	16.05$\pm$0.04	&	-	&	-	&	-	&	Leavitt	\\
2021-04-14	&	59318.11	&	-	&	16.09$\pm$0.05	&	-	&	-	&	-	&	Leavitt	\\
2021-04-14	&	59318.41	&	-	&	16.22$\pm$0.04	&	-	&	-	&	-	&	Leavitt	\\
2021-04-14	&	59318.06	&	-	&	15.92$\pm$0.02	&	15.77$\pm$0.02	&	15.71$\pm$0.02	&	15.69$\pm$0.03	&	Apogee	\\
2021-04-14	&	59318.90	&	-	&	16.03$\pm$0.04	&	-	&	-	&	-	&	Leavitt	\\
2021-04-15	&	59319.14	&	-	&	15.95$\pm$0.03	&	-	&	-	&	-	&	ROS2	\\
2021-04-15	&	59319.33	&	-	&	16.18$\pm$0.05	&	-	&	-	&	-	&	Leavitt	\\
2021-04-15	&	59319.87	&	-	&	15.90$\pm$0.04	&	-	&	-	&	-	&	Leavitt	\\
2021-04-16	&	59320.10	&	16.62$\pm$0.02	&	-	&	-	&	-	&	-	&	IO:O	\\
2021-04-16	&	59320.12	&	-	&	16.04$\pm$0.05	&	-	&	-	&	-	&	Leavitt	\\
2021-04-16	&	59320.95	&	16.71$\pm$0.06	&	15.88$\pm$0.02	&	15.75$\pm$0.02	&	15.76$\pm$0.02	&	-	&	Moravian	\\
2021-04-17	&	59321.98	&	-	&	16.10$\pm$0.05	&	-	&	-	&	-	&	Leavitt	\\
2021-04-19	&	59323.11	&	-	&	16.23$\pm$0.05	&	-	&	-	&	-	&	Leavitt	\\
2021-04-19	&	59323.14	&	-	&	15.96$\pm$0.05	&	15.84$\pm$0.03	&	15.81$\pm$0.05	&	15.78$\pm$0.09	&	ROS2	\\
2021-04-19	&	59323.98	&	17.30$\pm$0.08	&	16.01$\pm$0.02	&	15.84$\pm$0.02	&	15.82$\pm$0.03	&	-	&	Moravian	\\
2021-04-19	&	59323.99	&	-	&	16.24$\pm$0.05	&	-	&	-	&	-	&	Leavitt	\\
2021-04-20	&	59324.91	&	-	&	16.17$\pm$0.04	&	-	&	-	&	-	&	Leavitt	\\
2021-04-21	&	59325.96	&	-	&	-	&	-	&	16.02$\pm$0.07	&	-	&	AFOSC	\\
2021-04-22	&	59326.06	&	-	&	16.37$\pm$0.06	&	-	&	-	&	-	&	Leavitt	\\
2021-04-22	&	59326.93	&	-	&	16.39$\pm$0.07	&	-	&	-	&	-	&	Leavitt	\\
\hline
\end{tabular}
\end{table*}
\begin{table*}
\centering
\contcaption{(Continued) Sloan $ugriz$ AB magnitudes of SN 2021foa.}
\begin{tabular}{cccccccl}
\hline
Date & MJD & $u$ & $g$ & $r$ & $i$ & $z$ & Instrument \\
\hline
2021-04-23	&	59327.14	&	-	&	-	&	16.11$\pm$0.03	&	16.00$\pm$0.04	&	-	&	ROS2	\\
2021-04-23	&	59327.93	&	-	&	16.36$\pm$0.06	&	-	&	-	&	-	&	Leavitt	\\
2021-04-24	&	59328.10	&	-	&	16.39$\pm$0.03	&	16.13$\pm$0.03	&	16.23$\pm$0.04	&	16.22$\pm$0.05	&	Apogee	\\
2021-04-25	&	59329.06	&	-	&	16.45$\pm$0.03	&	16.26$\pm$0.03	&	16.25$\pm$0.05	&	16.26$\pm$0.06	&	Apogee	\\
2021-04-30	&	59334.12	&	-	&	17.22$\pm$0.03	&	16.92$\pm$0.04	&	16.95$\pm$0.05	&	16.74$\pm$0.06	&	Apogee	\\
2021-05-04	&	59338.96	&	18.75$\pm$0.02	&	-	&	-	&	-	&	-	&	ALFOSC	\\
2021-05-05	&	59339.89	&	-	&	17.53$\pm$0.08	&	17.10$\pm$0.10	&	17.34$\pm$0.10	&	-	&	Moravian	\\
2021-05-06	&	59340.11	&	-	&	17.59$\pm$0.04	&	17.32$\pm$0.03	&	17.16$\pm$0.04	&	16.99$\pm$0.09	&	ROS2	\\
2021-05-07	&	59341.01	&	-	&	-	&	-	&	-	&	17.01$\pm$0.03	&	ALFOSC	\\
2021-05-08	&	59342.86	&	-	&	17.73$\pm$0.04	&	17.37$\pm$0.07	&	17.34$\pm$0.06	&	-	&	Moravian	\\
2021-05-11	&	59345.11	&	-	&	17.96$\pm$0.08	&	17.50$\pm$0.05	&	17.66$\pm$0.14	&	17.28$\pm$0.32	&	ROS2	\\
2021-05-12	&	59346.91	&	-	&	-	&	17.60$\pm$0.01	&	-	&	-	&	ALFOSC	\\
2021-05-14	&	59348.01	&	-	&	18.28$\pm$0.05	&	17.76$\pm$0.06	&	17.70$\pm$0.07	&	-	&	Apogee	\\
2021-05-16	&	59350.15	&	-	&	18.42$\pm$0.04	&	17.86$\pm$0.02	&	17.71$\pm$0.04	&	-	&	ROS2	\\
2021-05-16	&	59350.91	&	-	&	-	&	-	&	-	&	17.64$\pm$0.03	&	IO:O	\\
2021-05-17	&	59351.94	&	-	&	18.55$\pm$0.06	&	18.05$\pm$0.05	&	18.01$\pm$0.04	&	-	&	Moravian	\\
2021-05-22	&	59356.08	&	-	&	18.77$\pm$0.10	&	18.23$\pm$0.06	&	18.37$\pm$0.09	&	-	&	ROS2	\\
2021-05-26	&	59360.95	&	-	&	-	&	-	&	-	&	18.02$\pm$0.17	&	Apogee	\\
2021-06-05	&	59370.98	&	-	&	19.50$\pm$0.01	&	19.32$\pm$0.01	&	19.18$\pm$0.01	&	18.83$\pm$0.01	&	ALFOSC	\\
2021-06-15	&	59380.93	&	-	&	-	&	19.46$\pm$0.04	&	-	&	-	&	OSIRIS	\\
2021-06-17	&	59382.91	&	-	&	-	&	-	&	19.48$\pm$0.04	&	19.13$\pm$0.04	&	ALFOSC	\\
2021-07-09	&	59404.93	&	-	&	-	&	19.55$\pm$0.02	&	19.41$\pm$0.04	&	18.81$\pm$0.06	&	ALFOSC	\\
\hline
\end{tabular}
\end{table*}

\begin{table}
\centering
\caption{ATLAS $c$ and $o$ AB magnitudes of SN~2021foa. The forced photometry is available at https://fallingstar-data.com/forcedphot/ }
\label{tab6}
\begin{tabular}{ccc}
\hline
Date & MJD & $c$ \\
\hline
2021-02-10 & 59255.51	&	20.36$\pm$1.06	\\	
2021-02-15 & 59260.52	&	19.65$\pm$0.50	\\	
2021-02-18 & 59263.43	&	19.26$\pm$1.11	\\	
2021-02-22 & 59267.52	&	18.84$\pm$0.31	\\	
2021-03-20 & 59293.61	&	15.40$\pm$0.02	\\	
2021-03-22 & 59295.45	&	15.36$\pm$0.02	\\	
2021-04-04 & 59308.46	&	15.32$\pm$0.03	\\	
2021-04-07 & 59311.39	&	15.51$\pm$0.04	\\	
2021-04-11 & 59315.41	&	15.84$\pm$0.05	\\	
2021-04-15 & 59319.39	&	15.83$\pm$0.02	\\	
2021-04-17 & 59321.39	&	15.70$\pm$0.03	\\	
2021-05-03 & 59337.44	&	17.35$\pm$0.06	\\	
2021-05-10 & 59344.37	&	17.42$\pm$0.08	\\	
2021-05-15 & 59349.37	&	18.26$\pm$0.09	\\	
2021-05-17 & 59351.35	&	17.96$\pm$0.16	\\	
2021-06-02 & 59367.35	&	19.26$\pm$0.23	\\	
2021-06-06 & 59371.35	&	19.71$\pm$1.00	\\	
2021-06-08 & 59373.32	&	19.13$\pm$0.40	\\	
2021-06-14 & 59379.34	&	19.74$\pm$0.61	\\	
2021-07-01 & 59396.31	&	19.70$\pm$0.99	\\	
2021-07-14 & 59409.26	&	19.72$\pm$0.18	\\	
2021-07-16 & 59411.25	&	19.35$\pm$0.57	\\	
\hline
Date & MJD & $o$ \\
\hline
2021-02-10 & 59255.50	&	>20.1		\\
2021-02-22 & 59267.54	&	19.23$\pm$0.05	\\
2021-02-26 & 59271.46	&	19.02$\pm$0.53	\\
2021-02-27 & 59272.51	&	19.25$\pm$0.24	\\
2021-03-04 & 59277.59	&	19.00$\pm$0.08	\\
2021-03-06 & 59279.53	&	19.18$\pm$0.61	\\
2021-03-15 & 59288.46	&	15.99$\pm$0.05	\\
2021-03-18 & 59291.55	&	15.60$\pm$0.02	\\
2021-03-19 & 59292.54	&	15.48$\pm$0.03	\\
2021-03-20 & 59293.60	&	15.43$\pm$0.02	\\
2021-03-22 & 59295.46	&	15.33$\pm$0.03	\\
2021-03-26 & 59299.45	&	15.15$\pm$0.01	\\
2021-03-27 & 59300.38	&	15.17$\pm$0.03	\\
2021-03-31 & 59304.61	&	15.19$\pm$0.04	\\
2021-04-01 & 59305.55	&	15.18$\pm$0.04	\\
2021-04-03 & 59307.51	&	15.24$\pm$0.08	\\
2021-04-04 & 59308.45	&	15.33$\pm$0.02	\\
2021-04-17 & 59321.39	&	15.71$\pm$0.03	\\
2021-04-19 & 59323.39	&	15.79$\pm$0.01	\\
2021-04-20 & 59324.44	&	15.85$\pm$0.02	\\
2021-04-23 & 59327.44	&	16.06$\pm$0.02	\\
2021-04-24 & 59328.37	&	16.18$\pm$0.04	\\
2021-04-28 & 59332.46	&	16.72$\pm$0.05	\\
2021-04-29 & 59333.50	&	16.76$\pm$0.06	\\
2021-04-30 & 59334.44	&	16.85$\pm$0.11	\\
2021-05-01 & 59335.47	&	17.00$\pm$0.08	\\
2021-05-03 & 59337.43	&	17.30$\pm$0.07	\\
2021-05-10 & 59344.39	&	17.38$\pm$0.07	\\
2021-05-14 & 59348.37	&	17.76$\pm$0.06	\\
2021-05-17 & 59351.36	&	17.93$\pm$0.07	\\
2021-05-19 & 59353.38	&	18.14$\pm$0.03	\\
2021-05-21 & 59355.37	&	18.00$\pm$0.07	\\
2021-05-26 & 59360.41	&	18.67$\pm$0.16	\\
2021-05-28 & 59362.39	&	18.93$\pm$0.14	\\
2021-06-01 & 59366.38	&	19.23$\pm$0.23	\\
2021-06-02 & 59367.34	&	19.61$\pm$0.48	\\
\hline
\end{tabular}
\end{table}
\begin{table}
\centering
\contcaption{(Continued) ATLAS $c$ and $o$ AB magnitudes of SN 2021foa.}
\begin{tabular}{ccc}
\hline
Date & MJD & $o$ \\
\hline
2021-06-08 & 59373.33	&	19.13$\pm$0.18	\\
2021-06-12 & 59377.32	&	19.30$\pm$0.57	\\
2021-06-14 & 59379.32	&	19.32$\pm$0.37	\\
2021-06-26 & 59391.32	&	19.37$\pm$0.54	\\
2021-06-27 & 59392.33	&	19.69$\pm$0.34	\\
2021-06-28 & 59393.30	&	19.41$\pm$0.48	\\
2021-07-01 & 59396.32	&	19.47$\pm$0.46	\\
2021-07-08 & 59403.32	&	19.71$\pm$0.43	\\
2021-07-16 & 59411.26	&	18.91$\pm$0.66	\\
2021-07-26 & 59421.28	&	19.50$\pm$0.87	\\
2021-08-03 & 59429.27	&	19.67$\pm$0.92	\\
\hline
\end{tabular}
\end{table}

\begin{table}
\centering
\caption{NIR VEGA magnitudes of SN 2021foa. All measuremenets are from the REMIR instrument.}
\label{tab7}
\begin{tabular}{cccc}
\hline
Date & MJD & $J$ & $H$ \\
\hline
2021-03-23	&	59296.24	&	14.75$\pm$0.05	&	14.65$\pm$0.07	\\
2021-03-25	&	59298.24	&	14.74$\pm$0.05	&	14.61$\pm$0.05	\\
2021-03-29	&	59302.19	&	14.56$\pm$0.10	&	14.65$\pm$0.05	\\
2021-03-31	&	59304.18	&	14.53$\pm$0.04	&	14.62$\pm$0.05	\\
2021-04-02	&	59306.01	&	14.60$\pm$0.06	&	14.65$\pm$0.08	\\
2021-04-06	&	59310.06	&	14.52$\pm$0.06	&	14.52$\pm$0.08	\\
2021-04-08	&	59312.36	&	14.87$\pm$0.06	&	14.63$\pm$0.11	\\
2021-04-11	&	59315.28	&	14.89$\pm$0.07	&	14.84$\pm$0.05	\\
2021-04-15	&	59319.15	&	14.89$\pm$0.04	&	14.87$\pm$0.04	\\
2021-04-19	&	59323.15	&	14.93$\pm$0.07	&	14.80$\pm$0.05	\\
2021-04-23	&	59327.15	&	15.27$\pm$0.14	&	15.18$\pm$0.13	\\
2021-04-27	&	59331.15	&	15.73$\pm$0.19	&	15.37$\pm$0.19	\\
2021-05-01	&	59335.08	&	15.91$\pm$0.09	&	15.52$\pm$0.11	\\
2021-05-06	&	59340.11	&	16.23$\pm$0.07	&	15.99$\pm$0.10	\\
2021-05-11	&	59345.11	&	16.69$\pm$0.12	&	16.65$\pm$0.18	\\
2021-05-16	&	59350.15	&	16.86$\pm$0.15	&	- \\
\hline
\end{tabular}
\end{table}

\begin{table*}
\centering
\caption{Log of the spectroscopic observations of SN 2021foa. The phases are relative to the $V$-band maximum epoch (MJD~59301.8). The spectra will be uploaded to the WISEREP database at https://www.wiserep.org/.} 
\label{tab2}
\begin{tabular}{lllllll}
\hline
Date & MJD & Phase & Coverage & Resolution & Exposure & Telescope + Instrument + Grism \\
 & & (d) & (\AA) & (\AA) & (s) & \\
\hline
2021-03-17 & 59290.0 & $-$12 & 3800-9000 & 14  & -    & NOT+ALFOSC+gr4 \\
2021-03-18 & 59291.0 & $-$11 & 5000-9000 & 40  & 1200 & Copernico+AFOSC+VPH6 \\
2021-03-19 & 59292.1 & $-$10 & 3700-8550 & 6.0 & 1800 & NOT+ALFOSC+gr7/gr8 \\
2021-03-22 & 59295.0 & $-$7 & 3900-7100 & 14  & 1200 & Copernico+AFOSC+VPH7 \\
2021-03-26 & 59299.1 & $-$3 & 3600-9650 & 14  & 900  & NOT+ALFOSC+gr4 \\ 
2021-04-06 & 59310.1 & +8  & 3700-7100 & 6.5 & 1500 & NOT+ALFOSC+gr7 \\ 
2021-04-06 & 59310.9 & +9  & 5000-9000 & 22  & 900  & Copernico+AFOSC+VPH6 \\ 
2021-04-09 & 59314.0 & +12 & 3700-9000 & 18  & 900  & NOT+ALFOSC+gr4 \\
2021-04-14 & 59319.0 & +17 & 3700-9000 & 14  & 1200 & NOT+ALFOSC+gr4 \\ 
2021-04-27 & 59331.9 & +30 & 3650-9100 & 13  & 1500 & NOT+ALFOSC+gr4 \\
2021-05-05 & 59339.0 & +37 & 6150-7750 & 2.9 & 1200 & TNG+LRS+VHRR \\ 
2021-05-06 & 59340.0 & +38 & 3600-10300 & 10 & 600  & TNG+LRS+LRB/LRR \\
2021-05-12 & 59346.9 & +45 & 4000-7550 & 3.4 & 1080 & GTC+OSIRIS+R2000B/R2500R \\
2021-05-28 & 59362.9 & +61 & 3800-9650 & 14  & 2400 & NOT+ALFOSC+gr4 \\ 
2021-06-05 & 59370.9 & +69 & 3800-9650 & 14  & 3600 & NOT+ALFOSC+gr4 \\
2021-06-15 & 59380.9 & +79 & 3700-7850 & 6.9 & 1800 & GTC+OSIRIS+R1000B \\
\hline
\end{tabular}
\end{table*}

\end{appendix}

\end{document}